\documentclass[aps, pra, reprint, noshowpacs, superscriptaddress,nofootinbib,longbibliography,floatfix,
]{revtex4-2}
\usepackage{graphicx}
\usepackage{array}
\usepackage{url}
\usepackage{multirow}
\usepackage{soul}
\usepackage{xcolor}

\usepackage{mathtools,amssymb}
\usepackage{tikz}
\usepackage{tabularx}
\usepackage{enumitem}
\usepackage{censor}
\usepackage{svg}
\DeclareUnicodeCharacter{2009}{\,}

\newcolumntype{L}{>{\raggedright\arraybackslash}X}
\newcolumntype{C}{>{\centering\arraybackslash}X}
\newcolumntype{s}{>{\hsize=48mm}L}
\newcolumntype{a}{>{\hsize=10.5mm}C}

\newcommand{\note}[1]{}

\begin{document}
\title{Survey of physics reasoning on uncertainty concepts in experiments: an assessment of measurement uncertainty for introductory physics labs}

\author{Michael Vignal}
\affiliation{JILA, National Institute of Standards and Technology and the University of Colorado, Boulder, CO 80309, USA}
\affiliation{Department of Physics, University of Colorado, 390 UCB, Boulder, CO 80309, USA}

\author{Gayle Geschwind}
\affiliation{JILA, National Institute of Standards and Technology and the University of Colorado, Boulder, CO 80309, USA}
\affiliation{Department of Physics, University of Colorado, 390 UCB, Boulder, CO 80309, USA}

\author{Benjamin Pollard}
\affiliation{Department of Physics, Worcester Polytechnic Institute, 100 Institute Road, Worcester MA, 01609, USA}

\author{Rachel Henderson}
\affiliation{Department of Physics \& Astronomy and CREATE for STEM Institute, Michigan State University, East Lansing, Michigan 48824, USA}

\author{Marcos D. Caballero}
\affiliation{Department of Physics \& Astronomy and CREATE for STEM Institute, Michigan State University, East Lansing, Michigan 48824, USA}
\affiliation{Department of Computational Mathematics, Science, \& Engineering, Michigan State University, East Lansing, Michigan 48824, USA}
\affiliation{Department of Physics and Center for Computing in Science Education, University of Oslo, 0315 Oslo, Norway}

\author{H. J. Lewandowski}
\affiliation{JILA, National Institute of Standards and Technology and the University of Colorado, Boulder, CO 80309, USA}
\affiliation{Department of Physics, University of Colorado, 390 UCB, Boulder, CO 80309, USA}

\begin{abstract}
Measurement uncertainty is a critical feature of experimental research in the physical sciences, and the concepts and practices surrounding measurement uncertainty are important components of physics lab courses. However, there has not been a broadly applicable, research-based assessment tool that allows physics instructors to easily measure students' knowledge of measurement uncertainty concepts and practices. To address this need, we employed Evidence-Centered Design to create the Survey of Physics Reasoning on Uncertainty Concepts in Experiments (SPRUCE). SPRUCE is a pre-post assessment instrument intended for use in introductory (first- and second-year) physics lab courses to help instructors and researchers identify student strengths and challenges with measurement uncertainty. In this paper, we discuss the development of SPRUCE's assessment items guided by Evidence-Centered Design, focusing on how instructors' and researchers' assessment priorities were incorporated into the assessment items and how students' reasoning from pilot testing informed decisions around item answer options. We also present an example of some of the feedback an instructor would receive after implementing SPRUCE in a pre-post fashion, along with a brief discussion of how that feedback could be interpreted and acted upon.
\end{abstract}

\maketitle

\section{Introduction}

Measurement is a central component of experimental scientific research, as all experimental measurements have some uncertainty. Proper consideration and handling of measurement uncertainty (MU) is critical for appropriately interpreting measurements and making claims based on experimental data. While some techniques for determining and using MU can be quite sophisticated, it is still possible (and desirable~\cite{pollard_introductory_2021}) to teach basic MU techniques in introductory science labs. In experimental physics, MU informs comparisons of multiple measurements~\cite{pollard_impact_2020} or between measurements and values predicted by models~\cite{dounas-frazer_modelling_2018}, and so instruction around MU can help students better understand the nature of experimentation. This and other important features of MU has led to policies and recommendations to include MU in introductory science courses~\cite{noauthor_analyzing_2022,kozminski_aapt_2014}. As developing proficiency with MU practices becomes an even more important goal in undergraduate physics labs, it is critical to be able to assess the level to which students are reaching this goal. 

Educators often wish to evaluate student learning around important concepts and practices---often articulated as learning goals or learning objectives~\cite{anderson_taxonomy_2001}---in order to inform and improve their instruction. To help instructors determine if learning goals are being achieved, the physics education research community has often developed and employed research-based assessments instruments (RBAIs), which Madsen, McKagan, and Sayre define as ``an assessment that is developed based on research into student thinking for use by the wider...education community to provide a standardized assessment of teaching and learning''~\cite{madsen_resource_2017}. It is important to note that RBAIs provide researchers and instructors with opportunities to assess student learning across time, institutions, and curricular and pedagogical changes in order to inform and improve instruction; they are not intended to evaluate individual students for the purpose of assigning grades. 

Developers of RBAIs often employ a theoretical framework during assessment development, such as Evidence Centered Design (ECD)~\cite{mislevy_evidence-centered_2005}, the Three-Dimensional Learning Assessment Protocol~\cite{laverty_analysis_2018}, or the framework described by Adams and Wieman \cite{adams_development_2011}. Such frameworks ``facilitate communication, coherence, and efficiency in assessment design and task creation''~\cite{mislevy_evidence-centered_2005}, typically by outlining steps or stages of assessment development, including exploratory research, data collection, and item development through to assessment delivery, scoring, and validation. ECD, the framework used in this work, also provides a structure for establishing evidence-supported claims about student reasoning based on student responses on the assessment: these claims are grounded in evidentiary arguments (a major focus of this paper) and contribute to the validity of the assessment instrument.

Of the RBAIs employed in physics labs, several focus on measurement and MU, albeit to varying extents. The Physics Measurement Questionnaire (PMQ) has been fundamental in articulating the point-like and set-like reasoning paradigms~\cite{campbell_teaching_2005} and in measuring the success of course transformations aimed at helping students shift towards more set-like reasoning~\cite{pollard_impact_2018, lewandowski_student_2018, pollard_methodological_2020,pollard_impact_2020}; the Physics Lab Inventory of Critical Thinking (PLIC)~\cite{walsh_quantifying_2019} has been used to assess the effectiveness of a scaffold and fade approach to teaching critical thinking in a physics course~\cite{holmes_teaching_2015}; and the Concise Data Processing Assessment (CDPA)~\cite{day_development_2011} has been used to identify changes in student performance around MU~\cite{mcloughlin_development_2019} and to look at student performance across genders~\cite{day_gender_2016}. 

While each of these assessments deals with MU in some way, there is not currently a widely-administrable RBAI that focuses explicitly on MU in introductory (first- and second-year) physics laboratory courses. To address this gap in assessments, we have developed the Survey of Physics Reasoning on Uncertainty Concepts in Experiments (SPRUCE) using the assessment development framework of ECD~\cite{mislevy_evidence-centered_2005}. It is our hope that SPRUCE will help instructors and researchers identify and improve instruction around measurement uncertainty concepts and practices that are challenging for introductory physics lab students.

In this paper, we present SPRUCE's assessment questions (hereafter referred to as ``assessment items'' or simply ``items'') and discuss their development. The goals of this paper are to demonstrate:
\begin{enumerate}
    \item A need for a widely-administrable assessment of measurement uncertainty and how SPRUCE will satisfy that need,
    \item The assessment item development and refinement process, as guided by ECD;
    \item Examples of evidentiary arguments, formed from student reasoning, that support our ability to make claims about student knowledge based on student responses to the assessment items; and
    \item An example of feedback for instructors and how that feedback might be interpreted.
\end{enumerate}

We begin by discussing, in Sec.~\ref{background}, the need for a new MU RBAI and how the framework of ECD can facilitate the development of such an assessment. In Sec.~\ref{development}, we describe the first three layers of ECD (\textit{domain analysis}, \textit{domain model}, and \textit{conceptual assessment framework}) and how we gathered information and made decisions to support the development of assessment items and evidentiary arguments. The development and refinement of these items and arguments is discussed in Sec.~\ref{assessmentimplementation}. In Secs.~\ref{validity} and~\ref{feedback}, we briefly discuss components of validity and how instructors might interpret and use the results of the instrument. In the final section, Sec.~\ref{futurework}, we summarize the work discussed in this paper and provide information for instructors and researchers who may be interested in using SPRUCE.

Future papers will discuss details of SPRUCE scoring, statistical validation, and claims about student learning, which all require a discussion of SPRUCE's scoring scheme using a new scoring paradigm. As discussion of this paradigm is beyond this paper's scope of introducing SPRUCE, in this paper, these topics are discussed only briefly to highlight how they informed item development.

\section{Background}
\label{background}

Over the last 30 years, research-based assessment instruments (RBAIs) have been used in physics classrooms to probe areas of interest and import for physics education researchers and physics instructors. Particularly notable examples of RBAI use include Mazur's use of assessment questions from the Force Concept Inventory~\cite{hestenes_force_1992} to probe student understanding of Newton's third law in his introductory physics lecture at Harvard~\cite{mazur_farewell_2009}, which led him (and others) to rethink what instruction in a lecture setting should look like~\cite{fagen_peer_2002} and Eblen-Zayas' use of the Colorado Learning Attitudes about Science Survey for Experimental Physics (E-CLASS~\cite{zwickl_epistemology_2014}) in her advanced lab courses, where she found that introducing metacognitive activities in an open-ended lab course had a positive impact on student enthusiasm and confidence~\cite{eblen-zayas_impact_2016}.

More broadly, RBAIs have been developed and deployed in the areas of mechanics~\cite{hestenes_force_1992,thornton_assessing_1998}\note{fci,fmce}, electrostatics~\cite{maloney_surveying_2001,wilcox_coupled_2014}\note{csem,CUE}, quantum mechanics~\cite{sadaghiani_quantum_2015}\note{QMCA}, and thermodynamics~\cite{rainey_designing_2020}\note{u-step}. In addition, assessments have been used to probe quantitative reasoning~\cite{white_brahmia_physics_2021}\note{piql}, beliefs about physics and physics courses~\cite{adams_new_2006}\note{class}, experimental research and lab courses~\cite{zwickl_epistemology_2014}\note{e-class}, modeling~\cite{dounas-frazer_characterizing_2018, rios_using_2019}\note{MAPLE}, and concepts and practices used in laboratory courses~\cite{campbell_teaching_2005,walsh_quantifying_2019,day_development_2011}. These and other assessments can be found on the PhysPort website~\cite{noauthor_physport_nodate}, and many of them are also accessible on LASSO~\cite{noauthor_lasso_nodate}.

In this section, we provide more detail about RBAIs that probe student proficiency in working with measurement uncertainty (MU). We highlight the strengths of these existing assessments, while also (as stated in our first research goal) arguing that there is still a need for a new assessment specifically probing MU in introductory physics labs. We then present initial work on the Survey of Physics Reasoning on Uncertainty Concepts in Experiments (SPRUCE), which is our response to the need for a new MU assessment instrument, and discuss how the framework of Evidence-Centered Design (ECD)~\cite{mislevy_evidence-centered_2017} informed this work. 

\subsection{Research-Based Assessment Instruments in Physics Labs}
\label{lab RBAIs}

The following sections discuss three existing RBAIs that include some assessment of MU topics. While each of these RBAIs has contributed to our collective understanding of student reasoning around measurement uncertainty, they each have limitations that point to a need for a widely-administrable assessment of measurement uncertainty for introductory physics labs.

\subsubsection{Physics Measurement Questionnaire}

The Physics Measurement Questionnaire (PMQ) consists of multiple-choice and open-response items adapted from the Procedural and Conceptual Knowledge in Science (PACKS) Project~\cite{millar_investigating_1994} for use with students at the University of Cape Town, South Africa~\cite{campbell_teaching_2005}. These items present decisions that students might face in a lab course and ask students which option they agree with (in a multiple-choice format), and then ask them to explain their reasoning (in an open-response format). Validation of the PMQ consisted of student interviews to ``check students’ understanding of the questions and the interviewer’s interpretation of their responses'' and to ``confirm that the probes presented sufficient alternatives covering a wide enough range of possibilities''~\cite{campbell_teaching_2005}\note{p 18}.

One of the most important findings to come out of the PMQ was the articulation of the point and set paradigms for student reasoning. These paradigms classify many types of student reasoning as being either point-like, indicating students believe that quantities measured have a true value that can be obtained with a single, perfect measurement; set-like, a typically more expert-like view that measurements will always have uncertainty and that a true value (if it exists) can never be perfectly known; or something else, usually with elements of both point-like and set-like perspectives. 

Despite the successes of the PMQ in articulating this paradigm and helping to inform course transformations, the assessment has two large limitations: the PMQ covers only a narrow range of ideas related to MU (primarily around distributions of results from repeated measurements), and the assessment is open response and therefore laborious to score. This second limitation is compounded by variance in student responses observed at different institutions, sometimes requiring instructors and researchers to first modify the scoring scheme provided by the developers of the PMQ~\cite{pollard_impact_2020}. 

\subsubsection{Physics Lab Inventory of Critical Thinking}

The Physics Lab Inventory of Critical Thinking was developed by physics education researchers at Cornell University and Stanford University to ``assess how students critically evaluate experimental methods, data, and models''~\cite{noauthor_physport_nodate}. The developers of the PLIC conducted multiple rounds of interviews and full-course piloting with several hundred students, as well as distributed the instrument to experts, to establish various forms of validity of the instrument including construct and concurrent validity~\cite{walsh_quantifying_2019}. The PLIC is contextualized in a small number of experiments, about which students are asked multiple questions, and the assessment is administered in an online format. 

The PLIC has been used to evaluate a ``scaffold and fade approach'' to instruction around making comparisons between measurements, or between measurements and models, for students in an introductory physics lab course~\cite{holmes_teaching_2015}. This approach involves  structured, explicit focus on a concept or practice initially (the ``scaffold''), which then ``fades'' over the course of instruction as student proficiency develops. Students who received this scaffold and fade instruction around making comparisons were much more likely to think critically about their results and propose possible improvements to their experimental setup than were students who had taken the course the previous year and not received this instruction~\cite{holmes_teaching_2015}.


The PLIC was explicitly designed to assess critical thinking, which the authors define as ``the ways in which one uses data and evidence to make decisions about what to trust and what to do.'' The authors aim to assess critical thinking in a lab setting, and while this includes components of MU, MU is not the primary focus of the assessment~\cite{walsh_quantifying_2019}. 

\subsubsection{Concise Data Processing Assessment}

The Concise Data Processing Assessment (CDPA) was developed by researchers at the University of British Columbia (UBC) to assess student proficiency around MU (primarily related to error propagation) and data handling~\cite{day_development_2011}. It consists of multiple-choice questions and can be presented in a pre-post format so as to probe student learning in a course. The CDPA was developed to complement the learning goals of a ``rigorous'' introductory physics lab, and the researchers used full-class piloting and student interviews to refine the assessment items. The CDPA developers established validity with data from 12 faculty and 11 graduate students who completed the assessment. 

The CDPA has been employed to explore if improvements in student proficiency with MU had an impact on their scores on E-CLASS~\cite{mcloughlin_development_2019}. While there were not enough matched pre-and post-instruction data to make comparisons of improvement on these two assessments, no correlation was found between CDPA scores and E-CLASS pre-instruction scores. However, the CDPA was found to be able to measure shifts in student proficiency, specifically positive shifts around content that was emphasized in the courses and negative shifts in content that was not emphasized in instruction. This study was conducted with participants in their second- or third-year laboratory course at the University of Helsinki.


As stated above, the CDPA was developed to complement an intensive introductory physics lab, but even still, it is a challenging assessment: as part of assessment development, graduate students at UBC were administered the assessment and scored, on average, just over 50\%, with post-test scores for first-year students averaging less than 40\%. In the second study discussed above, second- and third-year physics majors showed no improvement in CDPA scores from the pre- to post- assessment (with an overall score of around 40\%). As such, the CDPA may not be appropriate for many introductory physics labs, as its difficulty may limit its ability to identify trends and provide usable feedback for instructors.

\subsection{Assessment Development Framework: Evidence Centered Design}
\label{ecd}

To help guide the development of SPRUCE, we employed the assessment development framework of Evidence Centered Design (ECD)~\cite{mislevy_evidence-centered_2005} to help us incorporate instructor priorities around MU into the assessment instrument and to support the gathering of evidence of student reasoning that informs our interpretation and evaluation of student responses to the assessment items. Throughout this paper, we refer to these explanations that link student reasoning to student item responses as \textit{evidentiary arguments}.

ECD consists of five layers to facilitate ``communication, coherence, and efficiency in assessment design and task creation''~\cite{mislevy_evidence-centered_2005}. We list and briefly summarize these layers below:

\begin{itemize}[itemsep=-.2ex]	
    \item \textit{Domain Analysis}: gather information on the topic to be assessed, including from current instructors.
    
    \item \textit{Domain Model}: organize \textit{domain analysis} data by writing narrative assessment arguments that describe proficiencies to be measured (which we do via assessment objectives~\cite{rainey_designing_2020,vignal_affordances_2022}), acceptable evidence of such proficiencies, and the methods for gathering this evidence.
   
    \item \textit{Conceptual Assessment Framework}: operationalize assessment arguments to determine appropriate assessment features and item formats.
   
    \item \textit{Assessment Implementation}: write then iteratively pilot and revise assessment items while establishing evidentiary arguments that link observable data (student responses) to targeted claims about student reasoning, which will eventually be quantified via a scoring scheme.
    
    \item \textit{Assessment Delivery}: finalized implementation of assessment, scoring scheme, and instructor reports.
   
\end{itemize}

The first layer of ECD, \textit{domain analysis}, is the topic of a previous paper~\cite{pollard_introductory_2021} and briefly summarized below. \textit{Domain model}, \textit{conceptual assessment framework}, and especially \textit{assessment implementation} constitute the bulk of the work presented here, with a strong emphasis on piloting and evidentiary arguments. Our development of the quantitative scoring scheme used with SPRUCE (part of \textit{assessment implementation)} and the fifth layer (\textit{assessment delivery}) are briefly discussed in this paper and are instead the focus of upcoming papers.

\section{SPRUCE Development}
\label{development}

\subsection{\textit{Domain Analysis}}

The first steps towards developing an RBAI on MU were presented in a previous paper~\cite{pollard_introductory_2021}. In that work, we conducted and analyzed interviews with 22 physics lab instructors at institutions that spanned a range of sizes, highest degrees offered, selectivity, and student body demographics. In these interviews, we sought to identify instructor priorities when it came to the teaching and learning of MU. These interviews were semi-structured in nature and typically lasted around one hour.

Preliminary coding of these interviews was done to identify which concepts and practices instructors described as priorities or aspirational priorities for their courses. Instructors also talked about challenges for students and for instruction, including dealing with ideas taught in high school that students need to unlearn or refine, which informed our decisions of what content to include (or not include) in SPRUCE.

\subsection{\textit{Domain Modeling}}
\label{domainmodeling}

After the \textit{domain analysis}, \textit{domain modeling} involves ``articulat[ing] the argument[s] that [connect] observations of students’ actions in various situations to inferences about what they know or can do''~\cite{mislevy_evidence-centered_2005}. These assessment arguments are narrative in structure and describe the concepts and practices (i.e., the constructs) to be assessed, how evidence of student proficiency with respect to those concepts and tasks might be gathered, and how the items will allow students to demonstrate such proficiencies. It is in this stage that specific instrument items begin to take shape, as ideas gathered in the \textit{domain analysis} are reexpressed in terms of specific tasks.

To more explicitly embody the assessment priorities of instructors, we expressed our assessment arguments in terms of assessment objectives~\cite{rainey_designing_2020}. Assessment objectives (AOs) are ``\textit{concise, specific articulations of measurable desired student performances regarding concepts and/or practices targeted by the assessment}''~\cite{vignal_affordances_2022}: essentially, the AOs are the instrument's constructs. AOs are similar in concept and grain size to learning objectives~\cite{anderson_taxonomy_2001}, but they are designed to ``span the space of feasible, testable outcomes'' of an assessment~\cite{rainey_designing_2020}. As discussed in~\cite{vignal_affordances_2022}, AOs also provide a number of additional benefits for assessment development beyond organization of ideas collected in the domain analysis.\footnote{In \cite{vignal_affordances_2022}, we described the articulation of assessment arguments as being part of the \textit{conceptual assessment framework}: we now believe that it more appropriately belongs in the \textit{domain model}.} 

For SPRUCE, our AOs emerged from the qualitative codes developed during the \textit{domain analysis} and from the list of concepts and practices noted as being important to experts that was developed in Ref.~\cite{pollard_introductory_2021} and from a survey of instructor priorities around MU. These AOs were iteratively refined during item development, piloting, and development of our evaluation scheme.

Ultimately, we identified four main areas of concepts and practices into which all of our AOs can be organized, and these categories and their AOs resemble the dimensions and concepts developed to model MU content in secondary science education~\cite{priemer_learning_2018}:
\begin{itemize}
    \item Sources of uncertainty: estimating the size of uncertainty and identifying ways to reduce it
    \item Handling of uncertainty: uncertainty propagation and significant digits
    \item Distributions and repeated measurements: mean, standard deviation, standard error, and the importance of taking multiple measurements.
    \item Modeling: comparisons between explicit externalized models and the data
\end{itemize}

Because the modeling category pertained primarily to explicit comparisons between externalized models and data, we determined that AOs in this category fell outside of the scope of this assessment instrument. However, there are still elements of modeling, as defined by the Experimental Modeling Framework~\cite{zwickl_incorporating_2014}, that remain as integral parts of the other categories. Additionally, as described in \cite{vignal_affordances_2022} and discussed further in Sec.~\ref{assessmentimplementation}, some individual AOs in the other categories were also removed because of difficulties in establishing clear evidence of student reasoning. The finalized AOs are presented in Table~\ref{AOs}.\footnote{Using the AOs outlined in Table~\ref{AOs}, we can describe the PMQ as covering S1 and most of Distributions and Repeated measurements (with the exception of D5 and D6), the PLIC as covering Sources of Uncertainty and Distributions and Repeated Measurements (again with the exceptions of D5 and D6), and the CDPA as focusing primarily on Handling of Uncertainty (as well as graphical representations of data).}

\begin{table}[t]
    \caption{Final SPRUCE assessment objectives, organized by assessment objective category.}
    \label{AOs}
    \begin{tabularx}{\linewidth}{lX}\hline\hline
\multicolumn{2}{c}{\textbf{Sources of Uncertainty}}\\
S1 & Estimate size of random/statistical uncertainty by considering instrument precision\\
S2 & Identify actions that might improve precision\\
S3 & Identify actions that might improve accuracy\\\hline
\multicolumn{2}{c}{\textbf{Handling of Uncertainty}}\\
H1 & Identify when to use fractional versus absolute uncertainty\\
H2 & Propagate uncertainties using formulas\\
H3 & Report results with uncertainties and correct significant digits\\
H4 & Use concepts of uncertainty propagation to identify the largest contributor to uncertainty in a calculated value\\\hline
\multicolumn{2}{c}{\textbf{Distributions and Repeated Measurements}}\\
D1 & Articulate why it is important to take several measurements during experimentation\\
D2 & Articulate that repeated measurements will give a distribution of results and not a single number\\
D3 & Calculate and report the mean of a distribution for the best estimate of the measurement\\
D4 & Articulate that the standard deviation is related to the width of the distribution of measurements\\
D5 & Report the standard error (standard deviation of the mean) for the uncertainty in the mean of a distribution\\
D6 & Calculate the standard error from the standard deviation\\
D7 & Determine if two measurements (with uncertainty) agree with each other
\\\hline\hline
    \end{tabularx}
\end{table}

In practice, rather than a strictly narrative structure, our assessment arguments included: a narrative description of the task that would be presented to students; the AOs the item would assess and which responses would constitute evidence of proficiency; and a paragraph describing the rationale for why the item is appropriate. In a sense, these assessment arguments represent a hypothesis regarding a claim that the assessment will be able to make: if we present task X to students and they provide response Y, then we can conclude Z about their knowledge and reasoning around a particular AO. The connection between student responses and student reasoning comes from evidentiary arguments, which are developed during \textit{assessment implementation} and described in Sec.~\ref{subsec evidentiary arguments}.

While the literature on ECD portrays a fairly linear progression from one layer to the next, we took a more iterative approach in which we revisited and revised our work in previous layers (including \textit{domain modeling}) as we worked on subsequent layers.

\subsection{\textit{Conceptual Assessment Framework}}
\label{assessmentobjectives}

The third layer of the ECD framework involves operationalizing the assessment arguments developed in the second layer to inform the development of assessment items. This process includes deciding on the format of the assessment and the individual items and selecting a scoring paradigm.

In order to ensure a compact survey and reduce the cognitive load on students, we contextualized all of SPRUCE's assessment items within four experiments (as opposed to each item being a unique experimental context). Initial experimental contexts aligned with contexts discussed by instructors in the \textit{domain analysis} and were refined as needed to support the establishment of evidentiary arguments. The four experiments are summarized in Table~\ref{tab:Experiments} and described in more detail in Sec.~\ref{changesfrompiloting}.

\begin{table}[t]
    \centering
    \caption{SPRUCE Experiment Descriptions}
    \label{tab:Experiments}
    \begin{tabularx}{\linewidth}{p{25mm} L}\hline\hline
        Experiment &
        Description\\\hline
        Cart Acceleration\newline (Experiment 1) & A cart is released from rest to roll down a ramp as part of an experiment to determine the acceleration of the cart. Students are asked about taking multiple measurements and to identify the source of greatest uncertainty in their calculation of the acceleration.\\
        Mug Density\newline (Experiment 2) & The density of a mug is to be computed by measuring its mass and volume. Students are asked to identify uncertainties in each measurement and then propagate those uncertainties.\\
        Spring Constant\newline (Experiment 3) & A mass hangs from a spring and the period of (vertical) oscillation is used to determine a spring constant. Students are asked to estimate and propagate uncertainties and make comparisons between results.\\
        Breaking Mass\newline (Experiment 4) & Successive masses are added to a mass hanger until the string holding the mass hanger breaks. Students are asked to estimate uncertainty of a single measurement, make comparisons between results, and answer questions about taking many more measurements.\\\hline\hline
    \end{tabularx}
\end{table}

To develop an assessment that is easy to administer to a large number of students---twice, as SPRUCE is intended to be used pre-instruction and post-instruction---we opted for an online format for the assessment~\cite{van_dusen_online_2021,wilcox_alternative_2016} using the survey platform Qualtrics. We embedded digital calculators in all items in which students select or enter a numeric response, and we selected six potential item formats that facilitate automated evaluation.

The first three item formats are multiple choice (MC), multiple response (MR), and numeric open response (NOR). These formats contain a single prompt (or ``stem''~\cite{engelhardt_introduction_2009}) to which students respond by selecting a single answer (for MC items) or multiple answers (for MR items) from a list of answer options, or by entering a number into a text box (for NOR items). MC and MR items are the most common types of items on assessments, as they are straightforward to develop and evaluate. While NOR items were more complicated to evaluate, Qualtrics is able to exclude non-numeric responses from text boxes, meaning that student responses were sufficiently constrained that these items could be evaluating using a simple algorithm. 

The next three item formats involve the coupling of two questions: coupled multiple choice (CMC) items that have two coupled MC parts, coupled multiple response (CMR) that have a MC part followed by a MR part, and coupled numeric open response (CNOR) that have two NOR parts. These item types are examples of two-tier questions~\cite{treagust_development_1988} in which, rather than considering the answer options selected in either question independently, it is the combination of selections from the coupled questions that is evaluated.

For SPRUCE's CMR items, the multiple response answer options are \textit{reasoning elements} that allow students to compose a justification to their response to the multiple choice question, a design used in other physics assessments~\cite{wilcox_coupled_2014,pollard_creating_2020,rainey_validation_2022} that allows for evaluating complex student reasoning in a format that can be evaluated by a computer (as opposed to, for example, a free response justification that must be evaluated by a person or well-trained machine learning algorithm~\cite{wilson_classification_2022}). We used CNOR items to compare student values and uncertainties to see if students reported these quantities using appropriate significant digits, though as these items ask students to respond in a text box, student browsers may store student responses from the pre-instruction assessment and suggest or auto-fill them during the post-instruction assessment, and so quantities that factor into student responses on NOR and CNOR items are slightly different between pre- and post-instruction versions of the assessment. 

\subsection{A Brief Note on Scoring}

The selection of a scoring paradigm also impacts what types of items one might use in an assessment instrument. From the early stages of SPRUCE's development, we decided to have items relate to potentially more than one AO and to score each item once for each of the item's AOs. This approach resulted in the development of \textit{couplet scoring} in which a couplet is essentially an item viewed and scored through the lens of a single AO. As discussed in a paper under review as of this publication~\cite{vignal_couplet_2023}, the couplet becomes the unit of assessment for scoring, validation, and reporting student proficiencies, and it offers a number of affordances in these and other aspects of assessment development. For example, we found that couplet scoring scaffolded item development and refinement and helped us craft the questions that we wanted within the constraints of MC, MR, and NOR items.

\begin{figure}
    \centering
    \includegraphics[width=.45\textwidth]{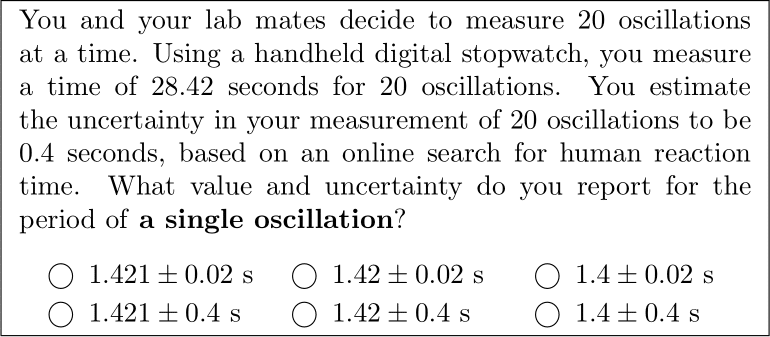}
    \caption{Item 3.3 (with modified numbers) asks students a single MC question about reporting the value of a single period of oscillation of a mass on a spring based on a measurement of 20 oscillations. The scoring of responses to this item is depicted in~Table \ref{tab33}.}
    \label{fig33}
\end{figure}

A simple example of couplet scoring for item~3.3 (Fig.~\ref{fig33}) is presented in Table~\ref{tab33}. In this item, students are asked what value they would report for the period of oscillation (with uncertainty) for a mass attached to a spring that is oscillating up and down. Students must select an answer based on information given in the prompt about a measurement of the time it takes the mass to complete 20 oscillations. Student responses are evaluated twice, once for ``H2 - Propagate uncertainties using formulas,'' and once for ``H3 - Report results with uncertainties and correct significant digits,'' as depicted in~\ref{tab33}. The independent scores from these couplets are not consolidated into a single item score (couplet scoring does not have ``item scores''), rather they each contribute (with all other couplets targeting the same AO) to independent AO scores, as discussed in Sec.~\ref{feedback}.

\begin{table}[t!]
    \centering
    \caption{Example scoring scheme for couplets of item~3.3. Student responses are scored once for each of the item's AOs ``H2 - Propagate uncertainties using formulas,'' and ``H3 - Report results with uncertainties and correct significant digits.''}
    \label{tab33}
    \begin{tabular}{llcc}\hline\hline
        \multicolumn{2}{l}{\multirow{2}{*}{Answer Option}} & \multicolumn{2}{c}{Score}\\
        & & H2 & H3                  \\\hline
        A & $1.421 \pm 0.02$ s  & 1 & 0 \\
        B & $1.421 \pm 0.4$ s   & 0 & 0 \\
        C & $1.42 \pm 0.02$ s   & 1 & 1 \\
        D & $1.42 \pm 0.4$ s    & 0 & 0 \\
        E & $1.4 \pm 0.02$ s    & 1 & 0 \\
        F & $1.4 \pm 0.4$ s     & 0 & 1 \\\hline\hline
    \end{tabular}
\end{table}

\section{Assessment Implementation}
\label{assessmentimplementation}

In the fourth layer of ECD, \textit{assessment implementation}, assessment items are written and, iteratively, pilot tested and refined as the developers construct the evidentiary arguments that facilitate meaningful interpretations of student responses. Items were constructed by expressing the assessment arguments developed in the \textit{domain analysis} in terms of the item formats identified in the \textit{conceptual assessment framework}.

\subsection{Evidentiary Arguments}
\label{subsec evidentiary arguments}
As stated above, the key focus of this paper, and a key component of ECD, is the establishment of evidentiary arguments, which allow researchers to map student reasoning to student responses. The primary source of evidence for evidentiary arguments in this work is student responses to the assessment items during pilot testing, though previous work with the PMQ,~\cite{pollard_impact_2020} and researcher expertise and experience also informed these arguments. 

Data from pilot testing (discussed in the next section) was used to establish our evidentiary arguments, linking student reasoning to student responses for each answer option, for each item, for each of the item's AOs. Interviews, especially, were used to probe student reasoning around not only students' final responses but also their `second best' responses and other responses they considered.

In an ideal situation, researchers would be able to make a one-to-one mapping between specific student responses and specific lines of student reasoning to ensure that evaluation is based on a perfectly accurate interpretation of student responses. In reality, no amount of piloting will capture all possible responses and reasoning employed by students, and so the goal is to develop evidentiary arguments to map trends in observed responses to trends in expressed reasoning. As a result, most of our item revisions were to improve our mappings by addressing instances in which different students either provided the same response with different justifications or provided different responses with the same justification.

\begin{figure}
    \centering
    \includegraphics[width=.45\textwidth]{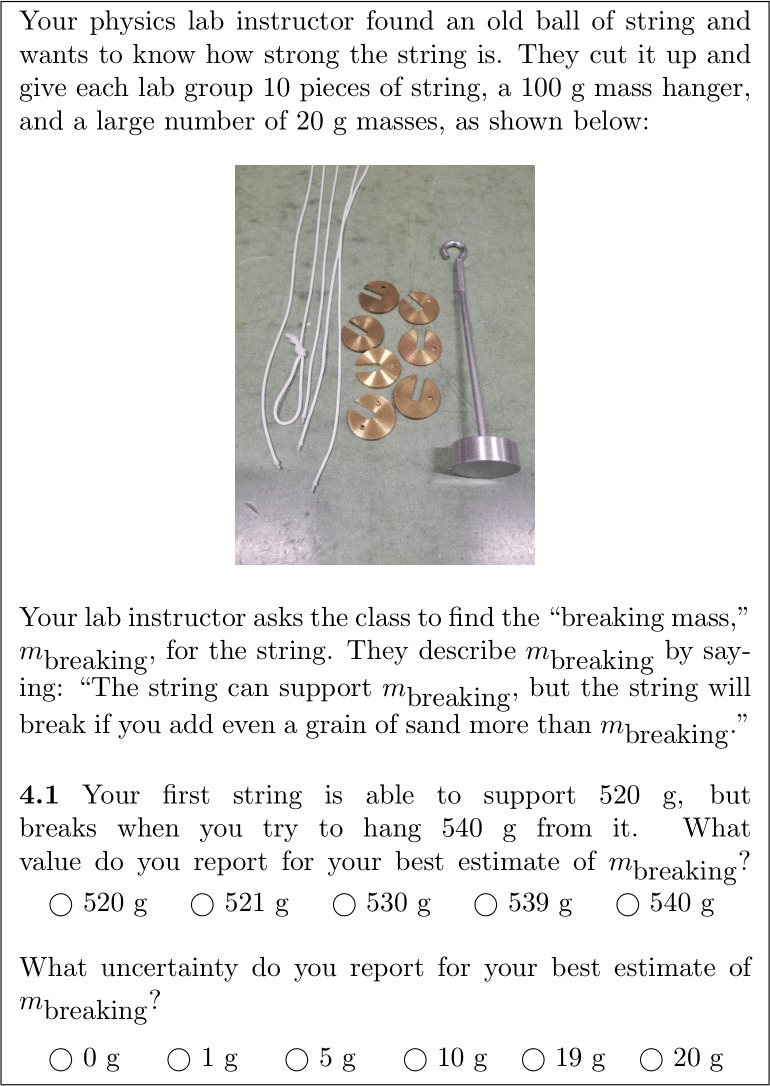}
    \caption{Item 4.1 went through iterations informed by multiple rounds of pilot testing with students.}
    \label{e4.1 and e4.2}
\end{figure}

To clearly illustrate what we mean by evidentiary arguments and how they were constructed and employed, we provide an example of our evidentiary arguments for item 4.1 (shown in Fig~\ref{e4.1 and e4.2}) in Table~\ref{Evidentiary Arguments}. This item is in a CMC (coupled multiple choice) format and only has one AO: ``S1 - Estimate size of random/statistical uncertainty by considering instrument precision.''

\begin{table*}[t]
     \centering
     \caption{Example Evidentiary Arguments for item~4.1. The top and bottom halves of the table includes the evidentiary arguments for the MC questions asking students, respectively, what mass and uncertainty they would report.}
     \label{Evidentiary Arguments}
     \begin{tabularx}{\linewidth}{csL}\hline\hline
     Answer Option  & Evidence-supported Reasoning
        & Example of Evidence \\\hline
     520 g          & Maximum Confirmed Supported Mass
        & ``520 is the last value reported that this string is able to support before it breaks.... So that's the closest value [to the breaking value] that we get before it breaks''\\ 
     521 g          & `Just Over' Maximum Confirmed Supported Mass
        & ``I guess it would be 521, since that wouldn't be too far [off from 520].''\\ 
     530 g          & Midpoint of 520 g and 540 g (often justified in conjunction with an uncertainty of 10 g)
        & ``We do know it's within the range of 520 to 540, and so what this does, if we have it at 530 with an uncertainty of 10, means our minimum value is just over 520, and maximum value is just under 540.'' \\ 
     539 g          & `Just Under' Minimum Confirmed Unsupported Mass
        & ``Maybe it's 539, because it breaks when you hit 540- maybe that was just slightly too big.'' \\ 
     540 g          & Minimum Confirmed Unsupported Mass
        & ``That's the value that the string broke on''\\\hline 
     0 g            & There is no uncertainty
        & Common Beta Response (not seen in interviews) \\
     1 g            & Small but non-zero uncertainty
        & ``It's better to include some uncertainty than to just make assumptions. So it wouldn't be zero, but it shouldn't be too far off.''\\ 
     5 g            & Half of Measurement Increment's `Place'
        & ``Like I said earlier, if it gives me one decimal place, my uncertainty would be the next one, like 05, so it can go up or down.''\\ 
     10 g           & Half of Measurement Increment
        & ``I picked 10 because the smallest increments that we can go in this measurement tool is 20, so I took the 20, divided by 2, and got 10''\\ 
     19 g           & Non-Inclusively Spans Range (e.g., 521 g to 540 g)
        & ``I would say 19 g...since it wouldn't include 520 but it could be anywhere else in that range [of 520] to 540.''\\ 
     20 g           & Measurement Increment
        & ``So I said 20 because we don't know what the- say like 521, 535, or 539, if that would also break. So there's uncertainty there, which I found because 540 - 520 is equal to 20.''\\\hline\hline 
     \end{tabularx}
\end{table*}

As our planned evaluation scheme evaluates each item along potentially multiple AOs, we established these mappings for each AO relevant to each item. In a few instances, when a mapping could be made for one AO but not another, the item was retained and simply not evaluated along the AO for which we could not establish sufficient evidentiary arguments.

The following sections discuss the different stages of piloting and many of the specific changes made to items as we worked to established evidentiary arguments.

\subsection{Piloting}

We implemented six pilot versions of the assessment between January and November 2022. These pilots consisted of multiple rounds of interviews and classroom implementation (which we refer to as ``beta piloting'' or simply ``betas''). The primary goals of piloting were to ensure that our items are appropriately interpreted by students and to collect sufficient evidence of student reasoning such that we could form comprehensive evidentiary arguments.

While each assessment item was intended to be presented to students in a particular format (e.g., MC, CMR, etc.), during piloting, we often temporarily changed the response format to gather additional information about student reasoning and student responses. These formatting decisions, as well as other priorities of the various pilots, are described in Table~\ref{tab:piloting}.

Even with fairly robust evidentiary arguments (as exemplified in Table~\ref{Evidentiary Arguments}) resulting from 39 interviews and beta testing with around 2000 students, it is likely there are examples of student reasoning that we did not observe. However, we worked to minimize such occurrences by recruiting as many students as possible from different types of institutions and introductory physics courses (using a database of instructors previously constructed~\cite{pollard_introductory_2021} and since expanded upon). Additionally, as this assessment is intended to inform instruction at the classroom level, not assign grades or otherwise evaluate students at an individual level, the impact of this limitation is further reduced by reporting averages and aggregated data to instructors and researchers.

Appendix~\ref{pilotingresponses} contains information about these courses and numbers of student participants in Table~\ref{Piloting} and student demographics in Table~\ref{demo}.

\begin{table}[t]
    \centering
    \caption{The item formats, primary goals, and number of student participants (N) for each of the six pilots (presented in chronological order).}
    \label{tab:piloting}
    \begin{tabularx}{\linewidth}{lLc}\hline\hline
        Pilot
        & Purpose(s)
        & N \\\hline
        Interviews 1 \ \
        & (Primarily Open-Response items)
        \newline Check Item Clarity
        \newline Establish Evidentiary Arguments
        \newline Identify Potential Refinements
        & 9 \\\hline
        Beta 1
        & (Primarily Closed-Response Items)
        \newline Preliminary Validation
        \newline Identify Potential Refinements
        \newline Pilot Scoring Scheme
        & 911 \\\hline
        Interviews 2 
        & (Primarily Closed-Response Items)
        \newline Check Item Clarity
        \newline Expand Evidentiary Arguments
        & 3 \\\hline
        Beta 2
        & (Primarily Open-Response Items)
        \newline Confirm MC Answer Options
        & 74 \\\hline
        Beta 3
        & (Primarily Final Item Formats)
        \newline Pilot Pre-instruction Implementation
        \newline Expand Evidentiary Arguments
        \newline Refine Scoring Scheme
        & 1048\\\hline
        Interviews 3
        & (Primarily Final Item Formats)
        \newline Finalize Evidentiary Arguments
        & 27
        \\\hline\hline
    \end{tabularx}
\end{table}

\subsubsection{Pilot Interviews}

Pilot interviews took place at three distinct stages of SPRUCE's development. The primary goal of these interviews was to gather evidence of student reasoning in order to establish evidentiary arguments linking student reasoning to student item responses. 
\newline

\noindent\textbf{Interviews: Round 1}

The first round of interviews was conducted to ensure item clarity, identify potential item refinements, and to begin developing evidentiary arguments.

Through course instructors, we solicited interview participants who had completed a introductory physics lab with a MU component in the previous 12 months. Nine students from four institutions were interviewed between January and February of 2022. Interviews were conducted with students completing the assessment on their computer while screen-sharing with the interviewer via Zoom. Interviews lasted between 30 minutes and 1 hour and were video and audio recorded. Students were compensated for their time with an electronic gift card.

In the interviews, students worked through the assessment items while the interviewer observed their responses and prompted students to provide reasoning supporting their final responses as well as other answers they considered. The majority of items were presented to students in an open response format.
\newline

\noindent\textbf{Interviews: Round 2}

A second set of interviews was conducted between June and August of 2022 to verify that our item distractors were sufficiently tempting and to again ensure that items and answer options were clear and understandable to students. We also further expanded our body of evidence of student reasoning by explicitly prompting students to explain their reasoning for not only their response but also, on many items, for a ``second-best'' response as well.

Interviews were solicited, conducted, and compensated in the same way as the first round of interviews. Despite the low number of participants, these interviews provided valuable data about student reasoning, especially for items that we had changed or were considering changing.
\newline

\noindent\textbf{Interviews: Round 3}

A final set of interviews to finalize our evidentiary arguments was conducted in October and November of 2022. We solicited interviewees (through instructors) from courses that participated in beta 3 (discussed below), so the majority of these students had already taken a prior version of the assessment. Twenty-seven interviews took place with students from eight courses across four institutions. These data provided substantial evidence of student reasoning and also identified a few items where our assessment was not capturing student reasoning as intended, prompting us to make a few minor modifications, and, as the interviews progressed, we began to see very few new ideas being expressed, indicating that we had likely conducted a sufficient number of interviews. These interviews were conducted and compensated in the same way as the previous interviews. 

\subsubsection{Full-class Beta Piloting}

During the Spring, Summer, and Fall 2022 terms, we conducted three full-class beta pilots of the assessment, where instructors asked students to take the assessment (generally outside of class). We encouraged instructors to offer participation credit or extra credit to students who completed the assessment, and in most of the courses the instructors did so. The assessment took most students around 20 minutes to complete, and they typically had at least a week in which to complete it. Instructors were given a list of students who had completed the assessment but were not given any information on individual student scores or responses.

For all three betas, students could complete the assessment for course credit (if awarded by the instructor) independent of if they consented to allow us to use their responses in our analysis, meaning students could complete the course assignment without granting us permission to use their responses in our analyses. We believe this this contributes to some of the courses having a rather low response rate as reported in Table~\ref{Piloting}, where we report the number and percentage only of students who consented to allowing us to use their responses in our research. Additionally, for betas 1 and 2, we removed students who did not complete at least two of the four experiments in the assessment, though by beta 3 (and in the final version of SPRUCE), we instead included a filter question (e.g., ``please enter the number 175 into the text box below'') at the end of the third experiment and removed students who did not reach the filter question or who answered it incorrectly. Filter questions have been used in previous assessments~\cite{wilcox_students_2017} to ensure the quality of responses that are analyzed for research, and unlike the system used in betas 1 and 2, they allow us to remove students who complete the assessment by selecting or entering random responses.

When applied to our data from beta testing, our scoring scheme allowed us to conduct \emph{preliminary} statistical validations of the instrument, specifically using classical test theory (CTT) with couplet scores (as opposed to item scores) as the unit of assessment~\cite{engelhardt_introduction_2009}. In instances where CTT indicated poorly performing couplets, we investigated the couplet to determine if and how to modify the item prompt, the answer options, and/or the scoring scheme. Several specific examples of these changes are given in Sec.~\ref{changesfrompiloting}, and a full CTT analysis is the focus of an upcoming paper. 
\newline

\noindent\textbf{Beta 1}

The first beta ran in the Spring of 2022, between interviews 1 and 2. This beta collected responses from students from eight courses at eight different institutions. In beta 1, almost all of the items were presented to students in a closed format (e.g., MC, MR as opposed to NOR), so that we could begin analyzing the distribution of students' responses across expected common response options, though for many items we did include a ``not listed'' option that allowed students to enter a response in a text box.

\begin{figure}[t]
    \centering
    \includegraphics[width=.45\textwidth]{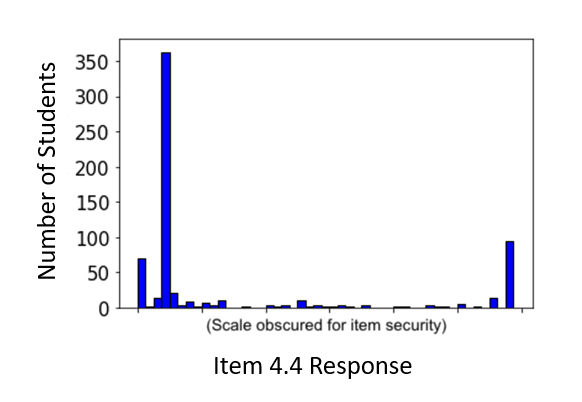}
    \caption{Histogram showing students' reported uncertainty values for item~4.4 on beta 1. These responses shows peaks at the correct value and at our planned distractors.}
    \label{fig:histogram_E4.4}
\end{figure}

However, one item, item 4.4, was presented in a NOR format despite being designed to be a MC item. Student responses to this item are shown in Fig.~\ref{fig:histogram_E4.4}. The distribution of student responses had peaks at values that corresponded to our planned answer options and, critically, there were no unexpected peaks indicating an attractive distractor that we had not anticipated. This finding informed the development of our second beta, discussed below.
\newline

\noindent\textbf{Beta 2}

The primary purpose of beta 2 was to verify the reasonableness of our distractors for MC items, and so items were presented to students primarily in an open-response format (e.g., NOR). While this beta was administered only to students in one course, the responses gathered strongly indicated that our previously identified distractors covered the most frequent incorrect answers provided by students, with only a few new distractors being identified through this beta. 
\newline

\noindent\textbf{Beta 3}\nopagebreak

The final round of piloting occurred during the beginning of the Fall 2022 term, where the assessment was administered prior to instruction (as SPRUCE is intended to be used in a pre-post modality). Items were primarily presented to students in their final format.

\subsection{Piloting-Informed Item Refinement}
\label{changesfrompiloting}

As discussed above, evidentiary arguments allow for a mapping between student reasoning and student item responses. When this is not the case, items should be modified or discarded. As our evaluation scheme considered each of an item's AOs independently, when modifying an item, we needed to consider each of the items AOs. In the following sections, we provide examples of how items were modified or removed based on our ability to develop sufficient evidentiary arguments. We do not discuss every evidentiary argument, item modification, or even every assessment item in these sections, rather we provide examples to represent the breadth of these arguments while highlighting the items for which establishing evidentiary arguments proved to be the most challenging. These sections are organized according to the four experiments that students work through on the assessment: brief descriptions of the four experiments are given in Table~\ref{tab:Experiments} and further detail is given in the following sections. In addition to changes informed by evidentiary arguments, many small formatting and wording changes informed by student interviews were made to ensure the items and answer options are clear and easily understood.

\subsubsection*{Experiment 0: Arrows on a Target}

The first item on SPRUCE is actually independent of the four experimental contexts and was added because of observed student difficulties during interviews in which students would consistently conflate accuracy and precision. The item presents four targets with different groupings of arrows and asks students to identify which grouping has high precision and low accuracy. This is a canonical scenario for discussing accuracy and precision in physics, and was added to allow for the possibility of `calibrating' our interpretation of student responses in items (specifically items~1.1, 3.2, and 4.8) that require students to distinguish between concepts of accuracy and precision in more complex scenarios.

While any such calibration would need to be supported by an empirical analysis of student responses, in theory, a student who conflated accuracy and precision on this item may still a have distinct, coherent, and largely correct understanding of these two concepts and may only be confusing the terms. Alternatively, this item may help identify if students who are able to correctly distinguish between accuracy and precision in this simple, likely familiar context are able to identify actions to improve accuracy and precision in more complicated, potentially unfamiliar situations.

\subsubsection*{Experiment 1: Cart Acceleration}

Experiment 1 presents students with an experiment to determine the acceleration of a cart rolling down a ramp. Specific assessment tasks are summarized in Table~\ref{tab:E1 Items}. 

\begin{table}[t]
    \centering
    \caption{Experiment 1 item types, descriptions, and AOs.}
    \label{tab:E1 Items}
    \begin{tabularx}{\linewidth}{ccLa}\hline\hline
        Item & Type & Description & AOs\\\hline
        1.1 & CMR & 
        Given a formula for acceleration, $a$, in terms of distance, $d$, and time, $t$, students are asked what they would do next (and why) after taking one measurement for $d$ and $t$. & S2, S3, D1, D2\\
        1.2 & MC & 
        Student are presented with values of $d$, $t$, and their uncertainties, and asked to reason about contributions to the uncertainty in the calculated value of $a$. & H1, H4\\\hline\hline
    \end{tabularx}
\end{table}

\begin{figure}[t]
    \centering
    \includegraphics[width=.45\textwidth]{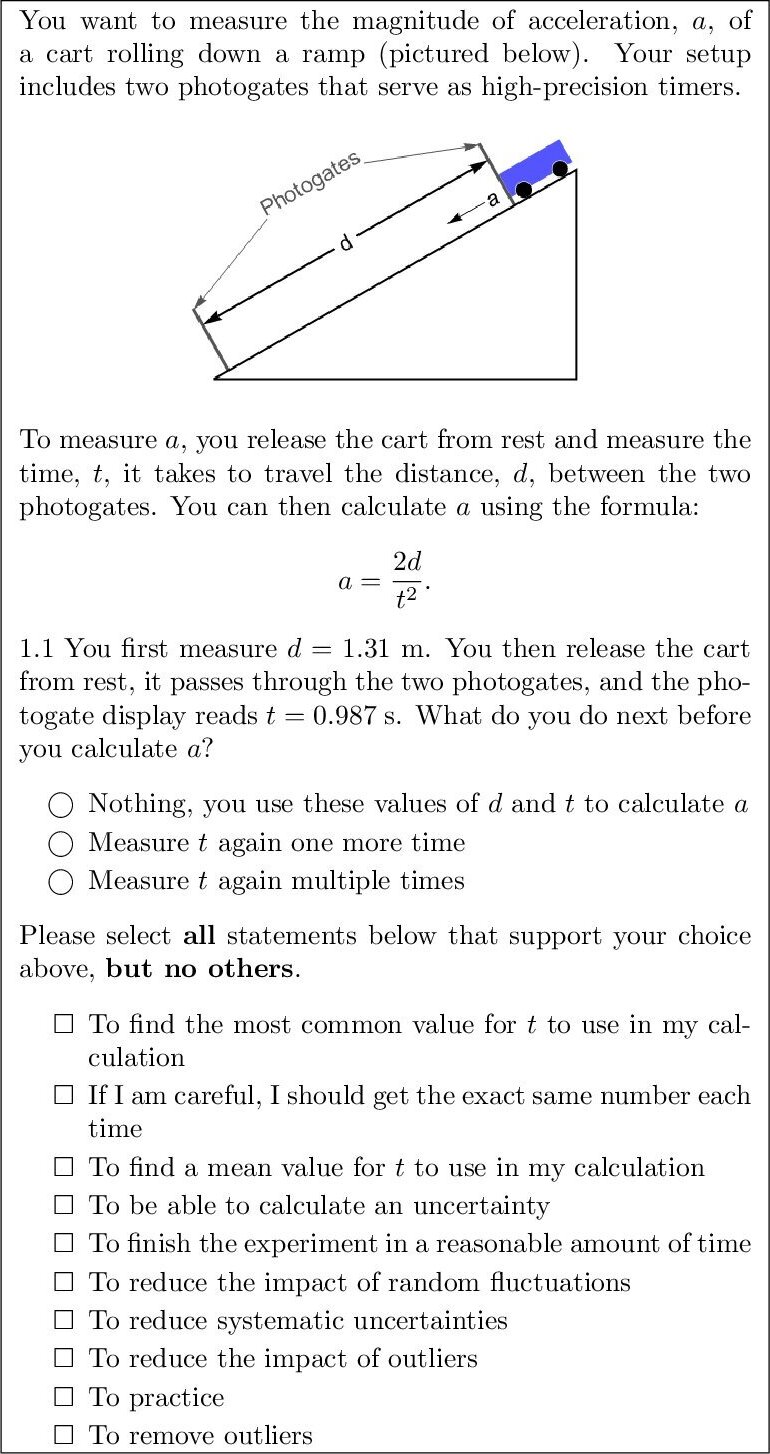}
    \caption{Item~1.1 asks students what they would do after taking a single measurement for time, then asks students to support that choice.}
    \label{e1.1}
\end{figure}

Item~1.1 (shown in Fig.~\ref{e1.1}) is largely modeled after the ``Repeating Distance'' item from the PMQ~\cite{campbell_teaching_2005}. Early iterations of this item consisted of MC and CMC questions, however the research team was unable to clearly establish evidentiary arguments because multiple explanations, some correct and some incorrect, would lead to different students selecting the same answer options. Eventually the team decided to present the item as a single CMR item (as shown in Fig.~\ref{e1.1}), in which students select an answer and also the reasoning that supports their answer. 

The reasoning elements in the MR part of the CMR item were initially derived from the codes used to score item RD on the PMQ~\cite{campbell_teaching_2005,pollard_impact_2020} and refined based on interviews 2 and 3 and beta 3. Care was taken to ensure that answer options were generally mechanistic in nature: for example, one of the early answer options, ``to improve accuracy,'' was removed because evidentiary arguments for this answer option were somewhat tautological, as the answer option was redundant with one of the item's AOs (``S3 - Identify actions that might improve accuracy)''. Instead, answer options that explained how accuracy could be improved were added to the item.

\subsubsection*{Experiment 2: Mug Density}

In experiment 2, students are asked to measure the mass and volume of a coffee mug to determine the mug's density (with uncertainty). Specific item tasks are summarized in Table~\ref{tab:E2 Items}.

\begin{table}[t]
    \centering
    \caption{Experiment 2 item types, descriptions, and AOs.}
    \label{tab:E2 Items}
    \begin{tabularx}{\linewidth}{ccLa}\hline\hline
        Item & Type & Description & AOs\\\hline
        2.1 & MC & Students are asked to report a value for the mass of the mug based on five measurements. & D3\\
        2.2 & MC & Students are asked to report an uncertainty for their value of the mass of the mug. & D5\\
        2.3 & CNOR & Students are shown before and after images of a graduated cylinder filled with water, where submerging the mug in the water has changed the level of the water line in the cylinders (and students are asked to report the values and uncertainties for the water levels before and after the mug is submerged). & S1,H3*\\
        2.4 & MC & Students are asked to propagate uncertainty in the water levels before and after submerging the mug through subtraction in order to determine the uncertainty in the measurement of the volume of the mug. & H1, H2\\
        2.5 & MC & Students are asked to propagate uncertainty in the mass and volume of the mug through division to determine the uncertainty in the calculated density of the mug.& H1, H2\\\hline\hline
    \end{tabularx}
    {\raggedright*H3 was eventually removed from item 3.2 due to our inability to establish clear evidentiary arguments.}
\end{table}

For item~2.2, interviews revealed that many students were selecting the correct answer of ``standard error (also known as standard deviation of the mean)'' because it contained the word ``mean'' (and most students had correctly calculated the mean in the previous item). However, when the parenthetical was removed for later interviews and betas, we observed that many students who knew the correct answer to be ``the standard deviation of the mean'' were unfamiliar with the term ``standard error.'' This presented the research team with a dilemma as both of these findings threatened our ability to confidently make evidentiary arguments for this item. Ultimately, we decided to keep the parenthetical to avoid arbitrarily large discrepancies between classes based on the particular language used in the course. This decision also impacted item~4.7, where we use the same language.


\subsubsection*{Experiment 3: Spring Constant}

In experiment 3, students are asked to determine the spring constant of a spring by first measuring the value of a mass and then the period of oscillation of that mass when it oscillates up and down while hanging from the spring. Specific task summaries are presented in Table~\ref{tab:E3 Items}.

\begin{table}[t]
    \centering
    \caption{Experiment 3 item types, descriptions, and AOs.}
    \label{tab:E3 Items}
    \begin{tabularx}{\linewidth}{ccLa}\hline\hline
        Item & Type & Description & AOs\\\hline
        3.1 & CNOR & Students are asked to identify the mass uncertainty in a single digital scale measurement. & S1, H3\\
        3.2 & CMR & Students are asked how many measurements or trials, and then how many oscillations per trial, they would use to measure the period of oscillation for a mass hanging vertically from a spring. Follow-up questions ask for justifications. & Trials: S2, S3, D1, D2 Oscillations: S2, S3\\
        3.3 & MC & Students are asked how they would report a value and uncertainty for a single oscillation based on a measurement of 10 oscillations and a given uncertainty estimate. & H2, H3\\
        3.4 & MR & Students are asked to identify means and uncertainties from other groups (represented numerically) that agree with their mean and uncertainty. & D7\\\hline\hline
    \end{tabularx}
\end{table}


Item 3.2 asks students to select and then justify the number of trials, and the number of oscillations per trial, they would use to obtain a measurement of the period of oscillation. Interviews revealed that different students were employing the same reasoning (e.g., wanting to minimize how much the period changed throughout the measurement) to justify different answers, and conversely other students were using different, often opposing, reasoning to justify the same answer. The research team ultimately elected to present this item in a double CMR format, with one MR follow-up asking students to justify the number of trials and the other to justify the number of oscillations per trial. Student interview responses to this item and to item 1.1 (which targets the same AOs), as well as a qualitative coding of student responses to a ``justify your answer'' free-response follow-up question on beta 3, informed the development of CMR reasoning element answer options. This item, in its CMR format, was then piloted in the third round of interviews, in which interviewers asked targeted follow-up questions to understand why students did or did not select specific answer options.

Item~3.4 asks students to determine if their measured value with uncertainty (reported numerically) agrees with the measurements of other groups. This item is isomorphic to item~4.3, which presents the exact same relative relationships between measurements using graphs. There is an abundance of research in the physics education literature regarding the use of various or multiple representations in physics (e.g., ~\cite{dufresne_solving_1997,rosengrant_case_2006,kohl_patterns_2008,vignal_investigating_2022, geschwind_representational_2023}, as well as in other STEM fields and more generally~\cite{cox_supporting_1995,parnafes_relations_2004,hand_examining_2010}), and these items provide researchers and instructors an opportunity to observe the impact of representation on student reasoning around comparing data. 

\subsubsection*{Experiment 4: Breaking Mass}

Experiment 4 intentionally asks students to consider measurement uncertainty in a novel situation: determining how much mass one must hang from a string before the string breaks. This situation is presented in item~4.1 as shown in Fig.~\ref{e4.1 and e4.2}, and the specific experiment tasks are summarized in Table~\ref{tab:E4 Items}. This item was intended to be novel for students while still being tractable, allowing us to evaluate student proficiency with various AOs in a novel context.

\begin{table}[t]
    \centering
    \caption{Experiment 4 item types, descriptions, and AOs.}
    \label{tab:E4 Items}
    \begin{tabularx}{\linewidth}{ccLa}\hline\hline
        Item & Type & Description & AOs\\\hline
        4.1 & CMC & Students are asked to identify $m_{breaking}$ and the uncertainty for a single measurement. & S1\\
        4.2 & NOR & Students are asked to report a value for the breaking mass based on 10 measurements. & D3\\
        4.3 & MR & Students are asked to compare their value (and an uncertainty we provide for them, both represented graphically) with the value and uncertainties of other groups. & D7\\
        4.4 & MC & Students are asked to calculate the standard error given the mean, number of measurements, and standard deviation. & D6\\
        4.5 & CNOR & Given the mean, number of measurements, standard error, and standard deviation, students are asked to report their value and uncertainty with appropriate significant digits. & H3, D5\\
        4.6 & MC & Students are asked what the impact on the standard deviation would be when going from 200 to 1000 measurements. & D4\\
        4.7 & MC & Students are asked what the impact on the standard error would be when going from 200 to 1000 measurements. & S2\\
        4.8 & MC & Students are asked what the impact on accuracy and precision would be when going from 200 to 1000 measurements. & S2, S3\\
        \hline\hline
    \end{tabularx}
\end{table}

Item~4.1 (shown in Fig.~\ref{e4.1 and e4.2}) is a somewhat unusual question for an experimental setting in that the resolution of the measurement is quite large (20 g), even for introductory physics labs. During interviews, this feature revealed interesting insights into student reasoning, and led to the refinement of the prompts and the inclusion of 521 g and 539 g (as the string was able to hold 520 g but broke when an additional 20 g was added) for the mass estimate and 1 g and 19 g for the uncertainty estimate. The evidentiary arguments for this item are presented in detail in Table~\ref{Evidentiary Arguments}.

Item~4.2 (also shown in Fig.~\ref{e4.1 and e4.2}) was initially developed, in part, to address an AO of identifying and removing outliers, as one of measurements given was substantially different from the rest. However, fewer than 10\% of students removed the outlier in beta piloting, and in interviews, students described not removing the outlier for many different reasons, including that they did not notice the outlier, did not think it was enough of an outlier to justify removal, or thought it was a substantial outlier but did not feel comfortable removing it without being able to explain why it was an outlier. 
For these reasons, we removed this AO from this item (and from the assessment as a whole), but, as this was not the only AO addressed by this item, the item remained in the assessment.




\section{Designing for, and establishing evidence for, validity}
\label{validity}

A valid instrument is one that measures what it says it measures and produces scores that are meaningful measures of the content assessed~\cite{hemphill_measurement_1950,rovinelli_use_1977,nunnally_psychometric_1994,aera_standards_nodate,engelhardt_introduction_2009,day_development_2011,lindell_establishing_2013}. Considerations of validity were a primary focus of the development team and led us to use ECD and create AOs, which in turn guided every step of instrument development discussed in this paper. Table~\ref{validity table} details several types of validity along with design features that support developing a valid instrument. The table also outlines evidence for each of these types of validity, though establishing evidence for validity is the primary goal of an upcoming paper.

\begin{table*}[]
    \centering
    \caption{Several types of validity, including design features intended to support that type of validity and the evidence needed to show that SPRUCE has that type of validity.}
    \label{validity table}
    \begin{tabularx}{\textwidth}{>{\hsize=0.16\hsize}L>{\hsize=0.4\hsize}L>{\hsize=0.6\hsize}LL}\hline\hline
    Validity Type & Definition & Design Features to Support Validity & Evidence of Validity\\\hline
    Content Validity & The instrument measures the intended content domain. & AOs derived from instructor interviews.\newline AOs reviewed by instructors throughout development. &  Independent matching of items to AOs by two physics education researchers with experimental backgrounds: initial and final agreements with developers of 93\% and 99\%, respectively. To be further evaluated in upcoming validation paper using statistical methods, though preliminary statistical validations were performed on piloting data and used to guide item refinement.\\
    Face Validity & Items appear to measure their intended construct.\note{Engelhardt p 14}& Items were created and refined to align with specific AOs.&Established during piloting interviews.\newline Items were also reviewed by instructors at various stages of development.\\
    External Validity & Results are generalizable beyond piloting population. & Instructor interviews and student piloting drew from many different institutions, as shown in~\cite{pollard_introductory_2021} and Tables~\ref{Piloting} and \ref{demo}. & To be established in an upcoming validation paper comparing results between piloting institutions and other institutions.\\
    Criterion Validity & Scores correlate with other metrics.& \multicolumn{2}{l}{Not explored due to limitations in our institutional review board protocol.}\\\hline\hline
    \end{tabularx}
\end{table*}

As outlined in Table~\ref{validity table}, design decisions made throughout SPRUCE's development were intended to contribute to SPRUCE's content, face, and external validity. Many of these decisions center on our use of AOs and our extensive piloting.

The types of validity presented in Table~\ref{validity table} are primarily qualitative in nature. Preliminary quantitative evidence of validity was established through statistical analyses of student responses from pilot phase data using CTT. A full suite of statistical validation statistics using a broad range of student responses to the the final assessment will be presented in future work, with such analyses using couplets and couplet scores, rather than items and item scores, as the units of assessment. Such analyses will include CTT, factoring, differential functioning, and pre-post results (i.e., concurrent validity~\cite{walsh_quantifying_2019}, and, eventually, item response theory (IRT)~\cite{yang_item_2014} or multidimensional IRT~\cite{stewart_multidimensional_2018}.

\section{Instructor Reports}
\label{feedback}

\begin{figure*}
    \centering
    \includegraphics[width=\textwidth]{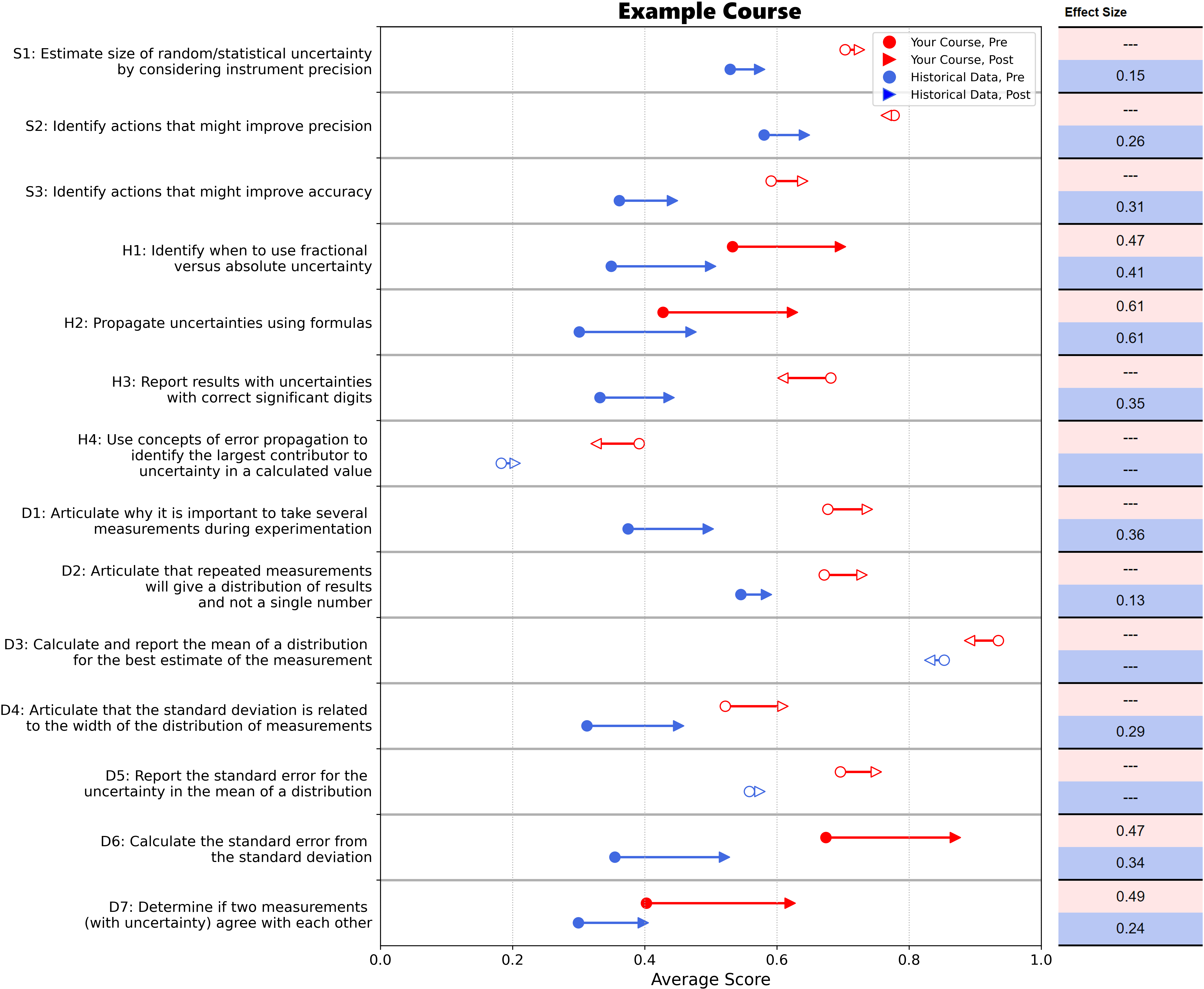}
    \caption{A portion of an instructor report showing pre- and post-instruction scores for the course and all historic data. Statistically significant shifts (as determined by a Mann-Whitney U test) for each AO are shown with solid circles and non-significant shifts are shown with open circles. Effect sizes for the statistically significant items are calculated using Cohen's d and shown on the right side of the chart.}
    \label{examplereport}
\end{figure*}

One of the main goals of developing an RBAI is to give instructors direct feedback about the impact of their course on student learning along the dimension measured by the assessment. For centrally administered RBAIs, instructors are often provided a report of the analysis of their students' performance. For SPRUCE, we provide an instructor report that not only provides the results from their students, but also comparison data from  all other courses that have used SPRUCE so far. The main graphic from such a report is shown in Fig.~\ref{examplereport}. The graph represents pre- and post-instruction scores for both the course and all historic data, with statistically significant shifts (as determined by a Mann-Whitney U test) for each AO shown with solid circles and non-significant shifts shown with open circles. Effect sizes for the statistically significant items are calculated using Cohen's d and shown on the right side of the chart. Because our data were not normal, which Cohen's D relies on for interpretation, we checked our findings using modified forms of Cohen's D~\cite{li_effect_2016}. This analysis produced similar qualitative effect sizes (small, medium, and large); thus, we report Cohen's D for simplicity. Not shown in Fig.~\ref{examplereport} are several paragraphs intended to support instructors in interpreting the graphic and effect size. These reports are based on the reports for the E-CLASS that were developed through interviews with instructors~\cite{wilcox_alternative_2016}, and will be refined as feedback from instructors whom implement SPRUCE continues to be collected.

For the course represented in this report, one can identify several important features. First, there are four AOs that show statically significant shifts from pre to post. Those include: (1) ``H1 - Identify when to use fractional verses absolute uncertainly,'' (2) ``H2 - Propagate uncertainties using formulas,'' (3) ``D7 - Determine if two measurements (with uncertainty) agree with each other, ''and (4) ``D6 - Calculate the standard error from the standard deviation.'' All of these shifts are positive (with post scores higher than pre scores) and have small effect sizes in the range of 0.11 \textendash 0.41. All four of these AOs align with course learning goals. However, there are many other AOs that also align with the course goals that show no statistically significant shifts. These observations can lead to actionable items for the course instructor. For instance, ``S3 - Identify actions that might improve accuracy'' is a main goal for the course, but shows no improvement over the semester. In this case, the instructor might consider interventions to target that goal, such as allowing for time in the lab for students to iteratively refine their apparatus, model, or data-taking procedure.

The second trend to note is that the students in this course score higher (even in the pretest) on average than students in all other courses combined. However, there larger gains (effect sizes) for many of the AOs for the historical data courses than for the target course. As we develop a large database of SPRUCE results, we can explore, as researchers, courses that succeed in having larger positive gains for specific AOs to understand possible casual effects using additional qualitative data. Additionally, we can explore many research questions using just the quantitative data. For example, we can determine correlations between the AO scores and the activities in the course (collected on the Course Information Survey) or demographic information collected. The results of these studies can then be used by instructors broadly as they make changes to improve their courses.

\section{Summary and Ongoing Work}
\label{futurework}

In this paper, we discussed the need for a widely-administrable assessment of measurement uncertainty for introductory physics labs and how we are using the assessment development framework of Evidence Centered Design (ECD)~\cite{mislevy_evidence-centered_2005} to create the Survey of Physics Reasoning on Uncertainty Concepts in Experiments (SPRUCE) to meet this need. While a previous paper~\cite{pollard_introductory_2021} discussed background research (\textit{domain analysis}), the layers of ECD discussed in this paper deal with: the creation of assessment objectives and assessment arguments (\textit{domain model}); instrument design, including the selection of a new scoring paradigm (\textit{conceptual assessment framework}); item development, piloting, and refinement with a focus on developing evidentiary arguments (\textit{assessment implementation}); and a portion of an example instructor report (\textit{assessment delivery}). Future papers will focus on additional aspects of \textit{assessment implementation} (e.g., scoring) and \textit{assessment delivery} (e.g., statistical evidence supporting validity).

While the ECD documentation depicts a fairly linear progression through the ECD layers, we found iteration across layers (outlined in Sec.~\ref{ecd}) to be extremely valuable and, in our view, necessary to gain the insights that informed the finalized products of the earlier layers. For example, item and AO development informed one another as we narrowed in on exactly what proficiencies we wanted to measure. Additionally, multiple rounds of piloting allowed us to present the same items to students using different formats (e.g., open response formats where students could input any answer, and closed response formats where students selected from a list of possible answer options), which allowed us to gather different types of data on student responses to create more robust evidentiary arguments. All together, these data informed our refinement of AOs, items formats, item prompts and answer options, and evidentiary arguments.




Instructors and researchers who are interested in using SPRUCE in their teaching and/or research can visit the SPRUCE website at~\cite{noauthor_spruce_nodate} for more information about how to use it in their own classes and studies.

\section{Acknowledgements}

This work is supported by NSF DUE 1914840, DUE 1913698, and PHY 1734006. We would also like to thank Robert Hobbs for his work contributing to the \textit{domain analysis}, as well as the instructors and students who contributed to the body of data upon which this assessment was built and refined.

\bibliography{bib.bib}
\clearpage
\onecolumngrid
\appendix

\section{Piloting Institutions, Responses, and Student Demographic Data}
\label{pilotingresponses}

\newcommand{\mc}[1]{\multicolumn{2}{C}{#1}}
\newcommand{\mr}[2]{\multirow{#1}{*}{#2}}

\begin{table*}[h]
     \centering     \caption{The number (N) and response rates (RR) of student participants in all six pilots, organized by course and institution. N is all of the students who consented to participate in the research study and who correctly answered a filter question located at the end of experiment 3: RR is this N value divided by the total number of students in the course. All courses were introductory laboratory courses at institutions in the US. R1 and R2 refer to Ph.D. granting institutions (with very high and high research intensity, respectively), M1 and M2 refer to master's granting institutions (with M1s being larger), BS and AS refer to bachelor's and associate's degree granting institutions (respectively), and MSI stands for minority serving institution.}
     \begin{tabularx}{\linewidth}{CCcCCCcccccc}\hline\hline
Institution & Institution   & Course    & Interview 1   & Interview 2   & Interview 3   & \mc{Beta 1}   & \mc{Beta 2}   & \mc{Beta 3}\\
Number      & Type          & Number    & N             & N             & N             & N     & RR    & N     & RR    & N & RR\\\hline
1           & R1            & 1         & 4             & 1             & 6             & 180   & 58\%  & 74    & 35 \% & 155   & 37 \%\\
2           & R2            & 2         & 3             & -             & -             & 123   & 40\%  & -     & -     & 218   & 31 \% \\
3           & R2            & 3         & 1             & -             & -             & -     & -     & -     & -     & -     & - \\
4           & R2            & 4         & 1             & -             & -             & 9     & 50\%  & -     & -     & -     & - \\
5           & R1            & 5         & -             & -             & 7             & 390   & 75\%  & -     & -     & 321   & 74\%\\
5           & R1            & 6         & -             & 2             & 8             & -     & -     & -     & -     & 112   & 91\%\\
6           & R1            & 7         & -             & -             & -             & -     & -     & -     & -     & 57    & 71\%\\
6           & R1            & 8         & -             & -             & 1             & -     & -     & -     & -     & 10    & 67\%\\
6           & R1            & 9         & -             & -             & 2             & -     & -     & -     & -     & 29    & 53\% \\
6           & R1            & 10        & -             & -             & 2             & -     & -     & -     & -     & 19    & 66\% \\
7           & AS            & 11        & -             & -             & 1             & -     & -     & -     & -     & 10    & 40\%\\
8           & R1            & 12        & -             & -             & -             & 128   & 31\%  & -     & -     & -     & - \\
9           & R1            & 13        & -             & -             & -             & 33    & 85\%  & -     & -     & 35    & 81 \%\\
10          & M2            & 14        & -             & -             & -             & 25    & 76\%  & -     & -     & -     & - \\
11          & M1, MSI       & 15        & -             & -             & -             & 23    & 71\%  & -     & -     & 17    & 35 \%\\
12          & AS            & 16        & -             & -             & -             & -     & -     & -     & -     & 54    & 93 \%\\
13          & R1            & 17        & -             & -             & -             & -     & -     & -     & -     & 20    & 95 \%\\ 
14          & BS/AS, MSI    & 18        & -             & -             & -             & -     & -     & -     & -     & 16    & 73 \%\\ 
15          & BS, MSI       & 19        & -             & -             & -             & -     & -     & -     & -     & 9     & 100 \%\\ 
16          & BS            & 20        & -             & -             & -             & -     & -     & -     & -     & 6     & 67 \%\\ 
 \hline\hline
     \end{tabularx}
     \label{Piloting}
\end{table*}

\begin{table*}[h]
     \centering
     \caption{Aggregate student demographics of students who participated in SPRUCE piloting and who elected to complete each of the optional demographic questions at the end of the survey.}
     \begin{tabular}{llcc}\hline\hline
\multicolumn{2}{l}{Demographic Category} & Interviews (N=39)   & Betas (N$\approx$1970)\\\hline
\multicolumn{2}{l}{Gender}\\
& Man                                           & 51\% & 59\%\\
& Woman                                         & 41\% & 39\%\\
& Non-Binary                                    & 8 \% & 2 \%\\
& Not Listed                                    & 0 \% & 1 \%\\\hline
\multicolumn{2}{l}{Race or Ethnicity}\\
& White                                         & 72\% & 75\%\\
& Asian                                         & 23\% & 16\%\\
& Hispanic/Latino                               & 10\% & 10\%\\
& Black or African American                     & 0 \% & 4 \%\\
& American Indian or Alaska Native              & 3 \% & 1 \%\\
& Native Hawaiian or other Pacific Islander     & 0 \% & 1 \%\\
& Not Listed                                    & 3 \% & 3 \%\\\hline
\multicolumn{2}{l}{English as a first language}\\
& Yes                                           & 87\% & 87\%\\
& No, but I am fluent in English                & 8 \% & 10 \%\\
& No, and I sometimes struggle with English     & 5 \% & 2 \%\\
& No, and I often struggle with English         & 0 \% & 1 \%\\
 \hline \hline
     \end{tabular}
     \label{demo}
\end{table*}
\end{document}